\documentclass[a4paper,11pt]{article}
\pdfoutput=1 

\usepackage{jcappub} 
\usepackage{indentfirst}
\usepackage{appendix}
\usepackage{multirow}

\newcommand{\be}{\begin{equation}}
\newcommand{\ee}{\end{equation}}
\newcommand{\bq}{\begin{eqnarray}}
\newcommand{\eq}{\end{eqnarray}}

\newcommand{\n}{\nonumber}
\def\({\left(}
\def\){\right)}
\def\[{\left[}
\def\]{\right]}

\title{Forecast for cosmological parameter estimation with gravitational-wave standard siren observation from the Cosmic Explorer}

\author[a]{Shang-Jie Jin,}
\author[a]{Dong-Ze He,}
\author[b,1]{Yidong Xu,}
\author[a]{Jing-Fei Zhang,}
\author[a,c,d,1]{Xin Zhang\note{Corresponding author.}}
\affiliation[a]{Department of Physics, College of Sciences, Northeastern
University, Shenyang 110819, China}
\affiliation[b]{Key Laboratory for Computational Astrophysics, National Astronomical Observatories, Chinese Academy of Sciences, Beijing 100101, China}
\affiliation[c]{Ministry of Education's Key Laboratory of Data Analytics and Optimization
for Smart Industry, Northeastern University, Shenyang 110819, China}
\affiliation[d]{Center for High Energy Physics, Peking University, Beijing 100080, China}

\emailAdd{850632821@qq.com, hedongze1992@163.com, xuyd@nao.cas.cn, jfzhang@mail.neu.edu.cn, zhangxin@mail.neu.edu.cn}

\abstract{The third-generation ground-based gravitational-wave (GW) detector, Cosmic Explorer (CE), is scheduled to start its observation in the 2030s. In this paper, we make a forecast for cosmological parameter estimation with gravitational-wave standard siren observation from the CE. We use the simulated GW standard siren data of CE to constrain the $\Lambda$CDM, $w$CDM and CPL models. We combine the simulated GW data with the current cosmological electromagnetic observations including the latest cosmic microwave background anisotropies data from Planck, the optical baryon acoustic oscillation measurements, and the type Ia supernovae observation (Pantheon compilation) to do the analysis. We find that the future standard siren observation from CE will improve the cosmological parameter estimation to a great extent, since the future GW standard siren data can well break the degeneracies generated by the optical observations between various cosmological parameters. We also find that the CE's constraining capability on the cosmological parameters is slightly better than that of the same-type GW detector, the Einstein Telescope. In addition, the synergy between the GW standard siren observation from CE and the 21 cm emission observation from SKA is also discussed.
}

\begin{document}
\maketitle

\section{Introduction}

Nowadays, the study of cosmology has entered the era of precision cosmology, and some great achievements have been made. For example, the accelerated expansion of the universe has been discovered \cite{Riess:1998cb,Perlmutter:1998np}; some cosmological parameters have been precisely measured \cite{Aghanim:2018eyx}; and in particular, a standard model of cosmology, which refers to the base six-parameter $\Lambda$ cold dark matter model (usually abbreviated as $\Lambda$CDM model), has been established. In this model, the cosmological constant (vacuum energy) $\Lambda$ with the equation of state (EoS) $w=-1$ serves as dark energy (DE) responsible for the late-time cosmic acceleration. The measurements of cosmic microwave background (CMB) anisotropies from Planck satellite mission \cite{Aghanim:2018eyx} have newly constrained the six primary parameters within the $\Lambda$CDM model with unprecedented precision. Notwithstanding, some parameter degeneracies still exist in the standard model. Thus, it is far-fetched to take the $\Lambda$CDM model as the eventual cosmic scenario. However, when the $\Lambda$CDM model is extended to include some new physics, e.g., a general dynamical dark energy, massive neutrinos, dark radiation, primordial gravitational waves, etc., the newly introduced parameters will severely degenerate with other parameters when the CMB alone data are used to constrain the extended models. Therefore, some other cosmological probes exploring the late-time universe are needed to be combined with the CMB data to break the parameter degeneracies.

The cosmological probes in light of the optical observations have been developed for many years. Besides the CMB observation, the mainstream cosmological probes mainly include, e.g., type Ia supernovae (SN), baryon acoustic oscillations (BAO), redshift space distortions (RSD), weak lensing, galaxy cluster counts, etc., which are all based on the observations of electromagnetic waves. In the future 10--15 years, these probes will be further greatly developed, and the fourth-generation DE programs, e.g., DESI \cite{Levi:2019ggs}, Euclid \cite{euclid}, and LSST \cite{lsst} will be implemented. However, other than those optical (or near-infrared) observations, some other new types of the cosmological probe should also be developed. The most promising new-type cosmological probes include the observations of gravitational waves (GWs) and of 21 cm radio waves.

The first detection of the GWs from the binary neutron star (BNS) merger (GW170817 \cite{TheLIGOScientific:2017qsa}; actually, the detection was only for the inspiral of the BNS) was rather meaningful, because it initiated the new era of multi-messenger astronomy. For this event, not only the GWs but also the electromagnetic waves in various bands \cite{GBM:2017lvd} were observed. It is well-known that GWs can serve as {\it standard sirens} since the waveform of GWs carries the information of luminosity distance to the source. Therefore, the multi-messenger observation of the GW standard sirens can provide us with both the information of distance and redshift, which can be used to study cosmology. For example, the event of GW170817 has been used to independently measure the Hubble constant \cite{Abbott:2017xzu}. The main advantage of such a standard siren method is that it avoids using the cosmic distance ladder. But due to the fact that we now have only one actual standard siren data point, the error of the Hubble constant measurement is still large (around 15\%), and thus it still cannot make an arbitration for the Hubble tension. Of course, with the accumulation of observed standard siren events, in the future the Hubble tension will be definitely resolved.

In the future, the GW standard sirens would also be developed into a powerful cosmological probe. The third-generation ground-based GW detectors have been proposed, e.g., the Einstein Telescope (ET) \cite{ET} in the Europe and the Cosmic Explorer (CE) \cite{Evans:2016mbw} in the United States. The detection ability of the third-generation ground-based GW detectors is much better than that of the second-generation ones \cite{Zhao:2017cbb,Zhao:2018gwk,Zhao:2017imr}. Using ET or CE, one can observe much more BNS merger events up to much higher redshifts, and thus they can play an important role in the cosmological parameter estimation and related issues in the future.

Recently, some issues concerning the forecasts for future cosmological parameter estimation based on the ET have been discussed \cite{Cai:2016sby,Li:2019ajo,Zhang:2018byx,Zhang:2019ylr,Zhang:2019ple,Zhang:2019loq,Wang:2018lun}. It is found that the GW standard siren observation from the ET would play a rather significant role in the measurement of cosmological parameters because the standard sirens can break the parameter degeneracies generated by the other cosmological observations (CMB combined with other lower-redshift observations). This is mainly because GW observation can provide a measurement of the absolute luminosity distance (not the relative one). In addition, the space-based GW observatories have also been proposed and are being built, e.g., the Laser Interferometer Space Antenna (LISA), TianQin and Taiji projects \cite{LISA,Mei:2015joa,Luo:2015ght,Hu:2018yqb,Wang:2019ryf,Shi:2019hqa,Wu:2018clg,Guo:2018npi,Tamanini:2016zlh}. The space-based GW detectors aim at detecting low-frequency GWs mainly generated from supermassive black hole coalescence and extreme mass ratio inspiral. The GW standard siren observations from the space-based GW observatories can also improve the cosmological parameter estimation to some extent \cite{Wang:2019tto,Zhao:2019gyk}, although there are still some uncertainties in the observations and data simulation.

In this paper, we will make a forecast for the cosmological parameter estimation using the GW standard siren observation from the CE. We will investigate what role the CE's GW standard siren observation would play in the future cosmological parameter measurement. We will focus the discussions on the cosmological parameter constraint capability of the CE, in particular for the cases considering the combination with the optical cosmological probes.
What's more, since both CE and ET belong to the third-generation ground-based GW detectors, in this work we also make a comparison for the promotion effects on the parameter estimation from CE and ET.

We consider three most typical DE cosmological models in this work, namely the $\Lambda$CDM model, the $w$CDM model, and the Chevalliear-Polarski-Linder (CPL) model. We will use the simulated standard siren data from the CE to constrain the cosmological parameters in these three DE models. Of course, the main task of this work is to investigate what role the standard sirens from the CE will play in combination with other cosmological observations. Therefore, we will combine the standard sirens with the current mainstream optical observations, i.e., the CMB+BAO+SN data, to investigate how the standard sirens from the CE would break the cosmological parameter degeneracies generated by the optical observations.

In addition to the GW standard sirens, the neutral hydrogen (HI) 21 cm emission observation is another important non-optical cosmological probe. In the future, the 21 cm radio observation by e.g. the Square Kilometre Array (SKA) \cite{SKA} will provide the HI power spectrum and the related BAO as well as RSD measurements, which will also play an important role in the future cosmological parameter estimation (see, e.g., the recent studies \cite{Bacon:2018dui,Xu:2016kwz,Xu:2014bya,Zhang:2019dyq,Zhang:2019ipd} and brief review \cite{Xu:2020uws}). Therefore, in this work, we will also discuss the synergy of the GW standard siren observation from the CE and the HI 21 cm emission observation from the SKA.

\section{Gravitational wave standard sirens}\label{sec2}


The most important advantage of the GW standard siren observation, compared with the current optical cosmological observations, is that the GW signals can provide the measurement of the absolute luminosity distance. Since the GW amplitude depends on the luminosity distance, we can naturally extract the information of luminosity distance from the analysis of the waveform.

For a Friedmann-Robertson-Walker universe, the line element reads
\begin{equation}
d{s^2} = -d{t^2} + {a^2}(t)\left[\frac{{d{r^2}}}{{1 - K{r^2}}} + {r^2}(d{\theta ^2} + {\sin ^2}\theta d{\phi ^2})\right],
\label{equa:ds}
\end{equation}
where $t$ is the cosmic time, $a(t)$ is the scale factor, and $K = +1$, $-1$, and $0$ correspond to closed, open, and flat universes, respectively. We set $G=c=1$ and $K = 0$ throughout this paper. In a flat universe, the luminosity distance $d_L$ can be written as
\begin{equation}
{d_L} (z)= \frac{{(1 + z)}}{{H_0}}\int_0^z {\frac{{dz'}}{{E(z')}}},
\label{equa:dl}
\end{equation}
where $E(z)\equiv H(z)/H_0$. 
For different DE models, the forms of $E(z)$ can be found in, e.g., ref.~\cite{Xu:2016grp}.

{Considering the transverse-traceless (TT) gauge, the strain $h(t)$ in the GW interferometers is given by \cite{Cai:2016sby}}
\begin{equation}
h(t)=F_+(\theta, \phi, \psi)h_+(t)+F_\times(\theta, \phi, \psi)h_\times(t),
\end{equation}
{where $F_{+, \times}$ are the antenna pattern functions, $\psi$ is the polarization angle, and ($\theta$, $\phi$) are angles describing the location of the source in the sky, relative to the detector. $h_{+}=h_{xx}=-h_{yy}$, and $h_{\times}=h_{xy}=h_{yx}$, which are the two independent components of the GW's tensor $h_{\alpha\beta}$ in the TT gauge.} Here, the antenna pattern functions of CE \cite{Sathyaprakash:2009xs} are
\begin{align}
F_+(\theta, \phi, \psi)=&~~\frac{1}{2}(1 + {\cos ^2}(\theta) )\cos (2\phi) \cos (2\psi)
                              - \cos (\theta) \sin (2\phi) \sin (2\psi),\n\\
F_\times(\theta, \phi, \psi)=&~~\frac{1}{2}(1 + {\cos ^2}(\theta) )\cos (2\phi) \sin (2\psi)
                              + \cos (\theta) \sin (2\phi) \cos (2\psi).
\label{equa:ce}
\end{align}

We also wish to make a direct comparison with the ET.
For ET, the antenna pattern functions \cite{Zhao:2010sz} become
\begin{align}
F_+^{(1)}(\theta, \phi, \psi)=&~~\frac{{\sqrt 3 }}{2}[\frac{1}{2}(1 + {\cos ^2}(\theta ))\cos (2\phi )\cos (2\psi )
                              - \cos (\theta )\sin (2\phi )\sin (2\psi )],\nonumber\\
F_\times^{(1)}(\theta, \phi, \psi)=&~~\frac{{\sqrt 3 }}{2}[\frac{1}{2}(1 + {\cos ^2}(\theta ))\cos (2\phi )\sin (2\psi )
                              + \cos (\theta )\sin (2\phi )\cos (2\psi )].
\label{equa:et}
\end{align}
{Since ET has three interferometers with $60^\circ$ inclined angles between each other, the other two antenna pattern functions are supposed to be} $F_{+,\times}^{(2)}(\theta, \phi, \psi)=F_{+,\times}^{(1)}(\theta, \phi+2\pi/3, \psi)$ and $F_{+,\times}^{(3)}(\theta, \phi, \psi)=F_{+,\times}^{(1)}(\theta, \phi+4\pi/3, \psi)$, respectively.

Following refs.~\cite{Zhao:2010sz,Li:2013lza}, we also apply the stationary phase approximation to the computation of the Fourier transform $\mathcal{H}(f)$ of the time domain waveform $h(t)$,
\begin{align}
\mathcal{H}(f)=\mathcal{A}f^{-7/6}\exp[i(2\pi ft_0-\pi/4+2\Psi(f/2)-\varphi_{(2.0)})],
\label{equa:hf}
\end{align}
where the Fourier amplitude $\mathcal{A}$ is given by
\begin{align}
\mathcal{A}=&~~\frac{1}{d_L}\sqrt{F_+^2(1+\cos^2(\iota))^2+4F_\times^2\cos^2(\iota)}
            \sqrt{5\pi/96}\pi^{-7/6}\mathcal{M}_c^{5/6},
\label{equa:A}
\end{align}
where $\mathcal{M}_c=M \eta^{3/5}$ is called ``chirp mass", $M=m_1+m_2$ is the total mass of coalescing binary with component masses $m_1$ and $m_2$, $\eta=m_1 m_2/M^2$ is the symmetric mass ratio. Here, it is necessary to point out that the observed chirp mass is related to the physical mass via $\mathcal{M}_{c,\rm obs}=(1+z)\mathcal{M}_{c,\rm phys}$~\cite{Cai:2016sby}. $\mathcal{M}_c$ in eq.~(\ref{equa:A}) represents the observed chirp mass. The constant $t_0$ denotes the time of the merger. $\iota$ is the angle of inclination of the binary's orbital angular momentum with the line of sight. The definitions of the functions $\Psi$ and $\varphi_{(2.0)}$ can refer to refs.~\cite{Zhao:2010sz,Li:2013lza}. Since it is expected that the short gamma ray bursts (SGRBs) are expected to be strongly beamed~\cite{Abdo:2009zza,Nakar:2005bs,Rezzolla:2011da}, the coincidence observations of SGRBs imply that the binaries are orientated nearly face on (i.e., $\iota\simeq 0$) and the maximal inclination is about $\iota=20^\circ$. Actually, averaging the Fisher matrix over the inclination $\iota$ and the polarization $\psi$ with the constraint $\iota<20^\circ$ is approximately the same as taking $\iota=0$ in the simulation~\cite{Li:2013lza}. Thus, we take $\iota=0$ in the calculation of the central values of the luminosity distances in the process of simulating GW sources.

\section{Method and data}\label{sec3}

\begin{figure*}[htbp]
\includegraphics[width=12cm,height=9cm]{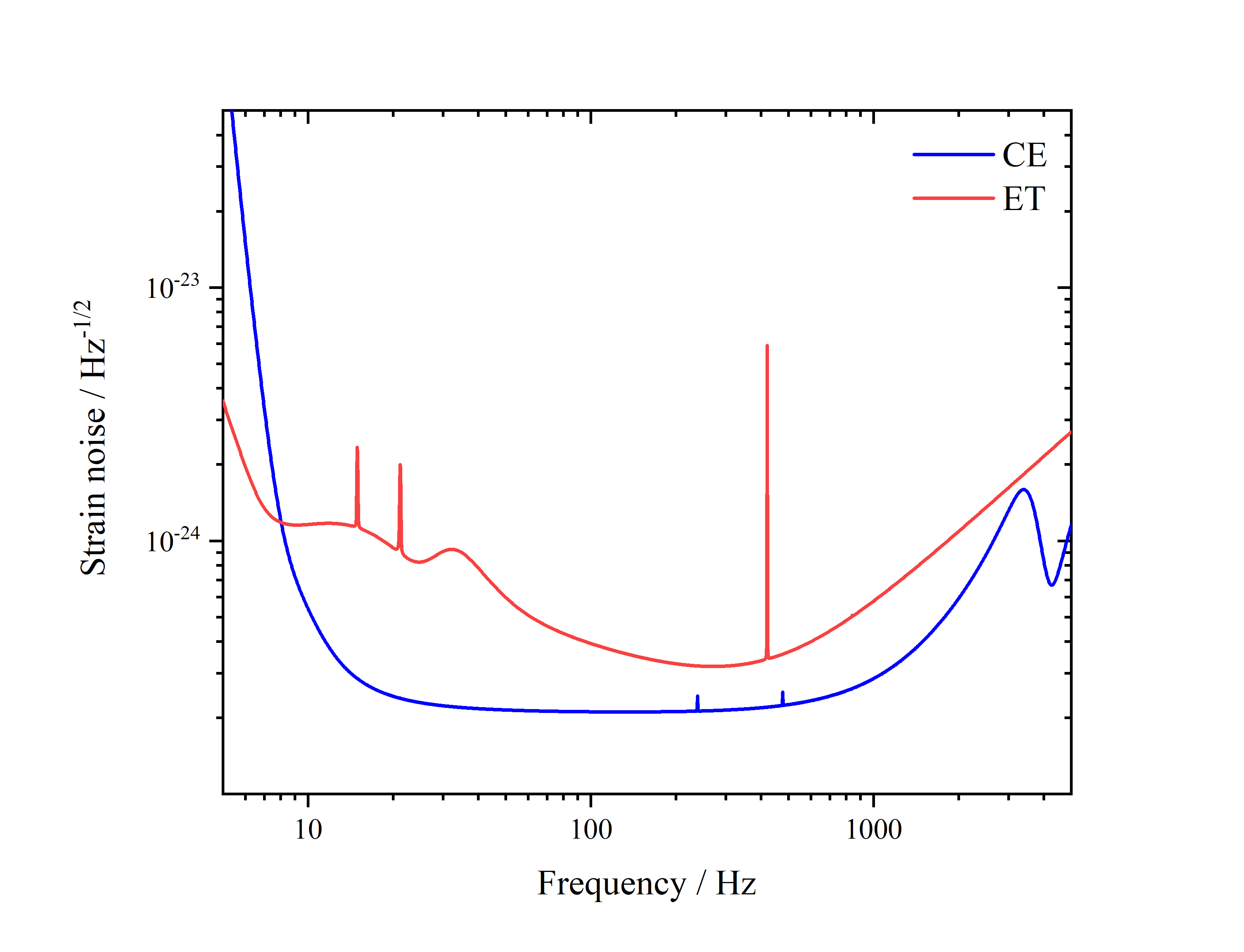}
\centering
{\caption{\label{fig:etce}The sensitivity curves of CE and ET, of which the text data files can be found in ref.~\cite{CEdata} and ref.~\cite{ET_Ddata}, respectively.}}
\end{figure*}

In this section, we shall introduce the method of simulating the GW standard siren data from the CE. In section~\ref{sec5}, since we discuss the comparison of CE and ET, we also introduce the method of simulating the GW standard siren data from the ET. We first describe how to get the errors of the luminosity distances to the GW sources in the simulation of standard siren data. Then, we introduce the current mainstream optical cosmological probes utilized in this work. Finally, we will use the Markov-chain Monte Carlo (MCMC) approach to infer the posterior distributions of cosmological parameters.


\subsection{Data simulation for the GW standard sirens}\label{sec3:a}

In this paper, we focus on the BNS systems and we assume that both neutron stars have mass of $m_1=m_2=1.4  M_{\odot}$, where $M_{\odot}$ is the solar mass. The first step for generating GW data is to simulate the redshift distribution of the sources. Following refs.~\cite{Cai:2016sby,Zhao:2010sz}, the redshift distribution of the sources takes the form
\begin{equation}
P(z)\propto \frac{4\pi d_C^2(z)R(z)}{H(z)(1+z)},
\label{equa:pz}
\end{equation}
where $d_C$ is the comoving distance, which is defined as $d_C(z)\equiv\int_0^z {1/H(z')dz'}$, and $R(z)$ describes the time evolution of the burst rate and takes the form~\cite{Schneider:2000sg,Cutler:2009qv}
\begin{equation}
R(z)=\begin{cases}
1+2z, & z\leq 1, \\
\frac{3}{4}(5-z), & 1<z<5, \\
0, & z\geq 5.
\end{cases}
\label{equa:rz}
\end{equation}

The combined signal-to-noise ratio (SNR) for the network of $N$ ($N=1$ for CE, and $N=3$ for ET) independent interferometers is given by
\begin{equation}
\rho=\sqrt{\sum\limits_{i=1}^{N}(\rho^{(i)})^2},
\label{euqa:rho}
\end{equation}
where $\rho^{(i)}=\sqrt{\left\langle \mathcal{H}^{(i)},\mathcal{H}^{(i)}\right\rangle}$. The inner product is defined as
\begin{equation}
\left\langle{a,b}\right\rangle=4\int_{f_{\rm lower}}^{f_{\rm upper}}\frac{\tilde a(f)\tilde b^\ast(f)+\tilde a^\ast(f)\tilde b(f)}{2}\frac{df}{S_h(f)},
\label{eq:product}
\end{equation}
where $\tilde a(f)$ and $\tilde b(f)$ are the Fourier transforms of the functions $a(t)$ and $b(t)$. Here, $f_{\rm lower}$ is the lower cutoff frequency. The upper cutoff frequency is dictated by the last stable orbit {(LSO) and it marks the end of the inspiral regime and the onset of the final merger.} Here, $f_{\rm upper}=2f_{\rm LSO}$, where $f_{\rm LSO}=1/{(6^{3/2}2\pi M_{\rm obs})}$ is the orbit frequency at the {LSO}, and $M_{\rm obs}=(1+z)M_{\rm phys}$ is the observed total mass~\cite{Zhao:2010sz}. The performance of a GW detector is characterized by the one-side noise \textit{power spectral density} (PSD) $S_h(f)$. {For the noise PSD of CE, we use the interpolation method to fit the CE stage 2 data in ref.~\cite{CEdata}.} While for ET, we take the noise PSD to be the same as described in ref.~\cite{Zhao:2010sz} and the SNR threshold is taken to be 8 for a detection.

Using the Fisher information matrix, we can get the instrumental error of $d_L$,
\begin{equation}
\sigma_{d_L}^{\rm inst}\simeq \sqrt{\left\langle\frac{\partial \mathcal H}{\partial d_L},\frac{\partial \mathcal H}{\partial d_L}\right\rangle^{-1}}.
\end{equation}
Because $\mathcal H \propto d_L^{-1}$, we can get $\sigma_{d_L}^{\rm inst}\simeq d_L/\rho$. When we simulate the GW sources, we take $\iota=0$ and add a factor 2 in front of the error~\cite{Li:2013lza}. More details can be found in ref.~\cite{Cai:2016sby}. {The instrumental error of $d_L$ can be written as}
\begin{equation}
\sigma_{d_L}^{\rm inst}\simeq \frac{2d_L}{\rho}.
\label{sigmainst}
\end{equation}
{Besides, the luminosity distance is affected by the weak lensing which will lead to an additional error of $d_L$~\cite{Zhao:2010sz,Sathyaprakash:2009xt},}
\begin{equation}
\sigma_{d_L}^{\rm lens}\simeq 0.05z d_L.
\end{equation}
Thus, the total error of $d_L$ is
\begin{align}
\sigma_{d_L}&~~=\sqrt{(\sigma_{d_L}^{\rm inst})^2+(\sigma_{d_L}^{\rm lens})^2} \nonumber\\
            &~~=\sqrt{\left(\frac{2d_L}{\rho}\right)^2+(0.05z d_L)^2}.
\label{sigmadl}
\end{align}
Now, we can simulate the measurement of the redshifts with the luminosity distances for the GW events of BNS mergers. More details can be found in ref.~\cite{Cai:2016sby}. Different from ref.~\cite{Zhang:2019loq}, we only consider the case of $1.4-1.4 M_{\odot}$ BNS merger.
Then we can generate the GW events with their $z$, $d_L$ and $\sigma_{d_L}$. In this work, we simulate 1000 GW standard siren events expected to be detected by the CE in the 2040s~\cite{Reitze:2019iox}.

\subsection{Current mainstream optical cosmological probes}

The current mainstream optical cosmological probes considered in this work include CMB, BAO, and SN. For the CMB data, we use the ``Planck distance priors" from the Planck 2018 results \cite{Aghanim:2018eyx,Chen:2018dbv}. For the BAO data, we use the measurements from 6dFGS at $z_{\rm eff}=0.106$ \cite{Beutler:2011hx}, SDSS-MGS at $z_{\rm eff}=0.15$ \cite{Ross:2014qpa}, and BOSS-DR12 at $z_{\rm eff}=0.38$, 0.51, and 0.61 \cite{Alam:2016hwk}. For the SN data, we use the latest sample from the Pantheon compilation \cite{Scolnic:2017caz}.
For convenience, the data combination ``CMB+BAO+SN" is also abbreviated as ``CBS" in the following.

\subsection{Combined constraints}

In order to constrain the cosmological parameters, we use the MCMC method to infer their posterior probability distributions. The total $\chi^2$ function of the combination of CMB, BAO and SN data is
\begin{equation}
\chi^2_{\rm tot} = \chi^2_{\rm CMB} + \chi^2 _{\rm BAO} +\chi^2_{\rm SN}.
\label{chi1}
\end{equation}

In this paper, we simulate 1000 GW data points. For the GW data, its $\chi^2$ can be written as
\begin{align}
\chi_{\rm GW}^2=\sum\limits_{i=1}^{1000}\left[\frac{\bar{d}_L^i-d_L(\bar{z}_i;\vec{\Omega})}{\bar{\sigma}_{d_L}^i}\right]^2,
\label{chi2}
\end{align}
where $\bar{z}_i$, $\bar{d}_L^i$, and $\bar{\sigma}_{d_L}^i$ are the $i$th redshift, luminosity distance, and error of luminosity distance, respectively. $\vec{\Omega}$ denotes a set of cosmological parameters.

If we consider the combination of the conventional cosmological electromagnetic observations and the GW standard siren observation, the total $\chi^2$ function becomes
\begin{equation}
\chi^2_{\rm tot} = \chi^2_{\rm CMB} + \chi^2 _{\rm BAO} +\chi^2_{\rm SN} +\chi^2_{\rm GW}.
\label{chi3}
\end{equation}

\begin{table*}[htbp]
\setlength\tabcolsep{1.0pt}
\renewcommand{\arraystretch}{1.5}
\centering
\resizebox{\textwidth}{!}
{
\begin{tabular}{cccccccccccccc}\\
\hline
\multicolumn{1}{c}{Model}&&\multicolumn{3}{c}{$\Lambda$CDM}&&\multicolumn{3}{c}{$w$CDM}&&\multicolumn{3}{c}{CPL}\cr
\cline{1-1}\cline{3-5}\cline{7-9}\cline{11-13}
Data&&CBS&GW&CBS+GW&&CBS&GW&CBS+GW&&CBS&GW&CBS+GW\cr
\hline
$\Omega_{\rm m}$&&$0.3142^{+0.0050}_{-0.0050}$&$0.3143^{+0.0062}_{-0.0062}$&$0.3142^{+0.0022}_{-0.0022}$&&
$0.3118^{+0.0077}_{-0.0077}$&$0.312^{+0.013}_{-0.012}$&$0.3115^{+0.0023}_{-0.0023}$&&$0.3120^{+0.0078}_{-0.0078}$&$0.296^{+0.065}_{-0.024}$&$0.3118^{+0.0043}_{-0.0043}$\cr
$H_0$&&$67.58^{+0.33}_{-0.33}$&$67.58^{+0.19}_{-0.19}$&$67.58^{+0.14}_{-0.14}$&&$67.92^{+0.83}_{-0.83}$&$67.93^{+0.34}_{-0.34}$&
$67.92^{+0.22}_{-0.22}$&&$67.92^{+0.84}_{-0.84}$&$67.87^{+0.71}_{-0.62}$&$67.93^{+0.37}_{-0.37}$\cr
$w$&&$-$&$-$&$-$&&$-1.015^{+0.032}_{-0.032}$&$-1.020^{+0.064}_{-0.058}$&$-1.014^{+0.017}_{-0.017}$&&$-$&$-$&$-$\cr
$w_0$&&$-$&$-$&$-$&&$-$&$-$&$-$&&$-0.992^{+0.083}_{-0.083}$&$-0.952^{+0.089}_{-0.120}$&$-0.993^{+0.050}_{-0.050}$\cr
$w_a$&&$-$&$-$&$-$&&$-$&$-$&$-$&&$-0.098^{+0.330}_{-0.280}$&$-0.19^{+1.20}_{-0.60}$&$-0.09^{+0.17}_{-0.15}$\cr
\hline
\end{tabular}
}
\caption{\label{tab1}The best-fit values for the parameters in the $\Lambda$CDM, $w$CDM and CPL models by using the CBS, GW, and CBS+GW data combinations. Here, CBS stands for CMB+BAO+SN and $H_0$ is in units of ${\rm km\ s^{-1}\ Mpc^{-1}}$.}
\end{table*}

\begin{table*}[htbp]
\small
\setlength\tabcolsep{2.8pt}
\renewcommand{\arraystretch}{1.5}
\centering
\begin{tabular}{ccccccccccccccc}
\\
\hline &\multicolumn{3}{c}{$\Lambda$CDM}&& \multicolumn{3}{c}{$w$CDM}&& \multicolumn{3}{c}{CPL}\\
 \cline{2-4}\cline{6-8}\cline{10-12}
  & CBS & GW &CBS+GW & &CBS & GW   &CBS+GW & &CBS & GW &CBS+GW\\
\hline
$\sigma(\Omega_{\rm m})$
                   & $0.0050$
                   & $0.0062$
                   & $0.0022$&
                   & $0.0077$
                   & $0.0125$
                   & $0.0023$&
                   & $0.0078$
                   & $0.0445$
                   & $0.0043$\\

$\sigma(H_0)$
                    & $0.330$
                    & $0.190$
                    & $0.140$&
                   & $0.830$
                   & $0.340$
                   & $0.220$&
                   & $0.840$
                   & $0.665$
                   & $0.370$\\

$\sigma(w)$
                    & $-$
                   & $-$
                   & $-$&
                   & $0.032$
                   & $0.061$
                   & $0.017$&
                 & $-$
                 & $-$
                 & $-$\\

$\sigma(w_0)$
                    & $-$
                   & $-$
                   & $-$&
                   & $-$
                 & $-$
                   & $-$&
                    & $0.0830$
                    & $0.1045$
                 & $0.0500$\\

$\sigma(w_a)$
                    & $-$
                   & $-$
                   & $-$&
                   & $-$
                 & $-$
                   & $-$&
                    & $0.305$
                    & $0.900$
                 & $0.160$\\\hline
$\varepsilon(\Omega_{\rm m})$
                   & $0.0159$
                    & $0.0197$
                    & $0.0070$&
                   & $0.0247$
                   & $0.0401$
                   & $0.0074$&
                   & $0.0250$
                   & $0.1503$
                   & $0.0138$\\

$\varepsilon(H_0)$
                    & $0.0049$
                    & $0.0028$
                    & $0.0021$&
                   & $0.0122$
                   & $0.0050$
                   & $0.0032$&
                   & $0.0124$
                   & $0.0098$
                   & $0.0054$\\

$\varepsilon(w)$
                    & $-$
                   & $-$
                   & $-$&
                   & $0.0315$
                   & $0.0598$
                   & $0.0168$&
                 & $-$
                 & $-$
                 & $-$\\

$\varepsilon(w_0)$
                    & $-$
                   & $-$
                   & $-$&
                   & $-$
                 & $-$
                   & $-$&
                   & $0.0837$
                   & $0.1098$
                 & $0.0504$\\
\hline
\end{tabular}
\centering
\caption{\label{tab2} Constraint errors and accuracies for the parameters in the $\Lambda$CDM, $w$CDM and CPL models by using the CBS, GW, and CBS+GW data combinations. Here, CBS stands for CMB+BAO+SN.}
\end{table*}

\begin{figure*}[htbp]
\includegraphics[width=7.2cm,height=5.4cm]{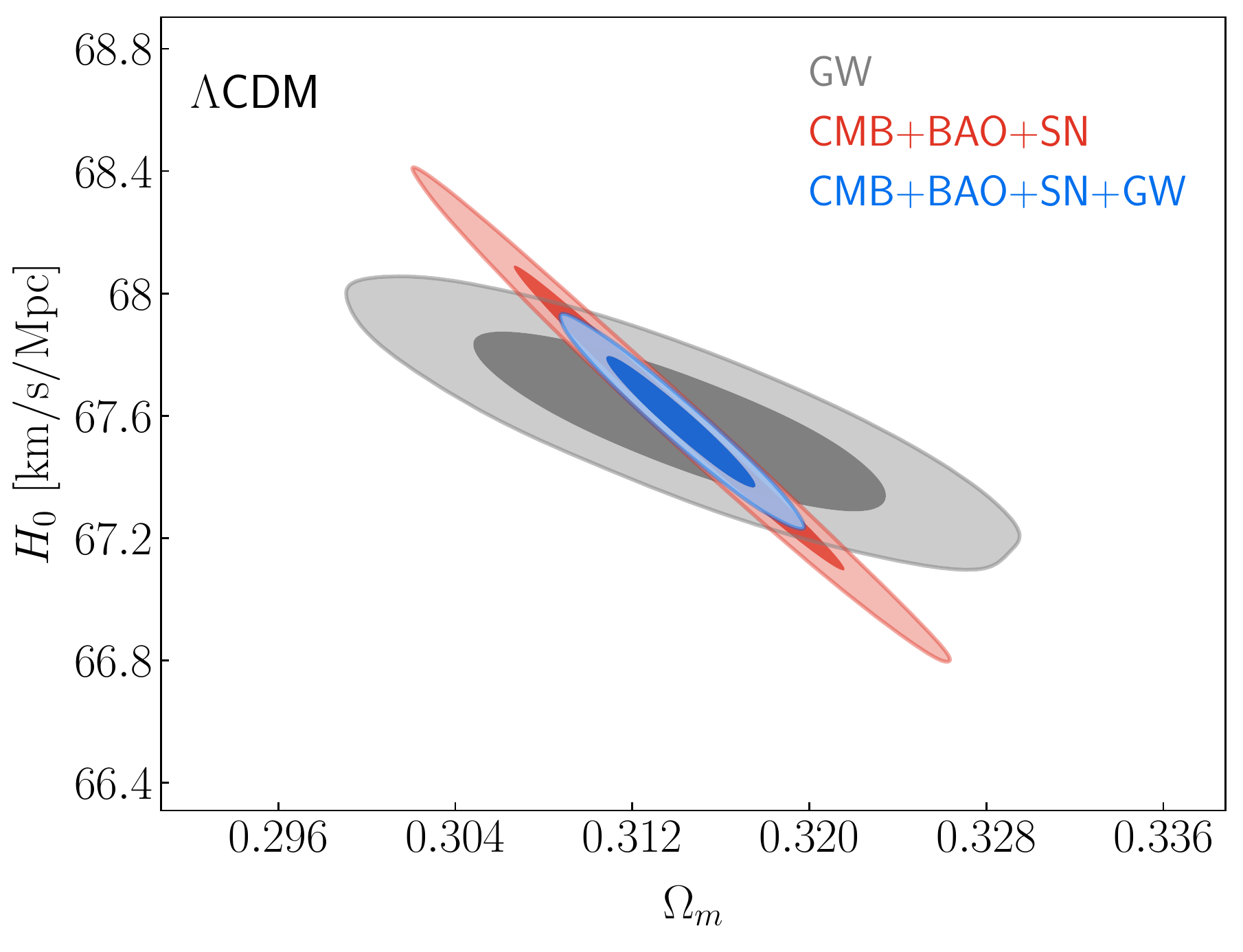}
\includegraphics[width=7.2cm,height=5.4cm]{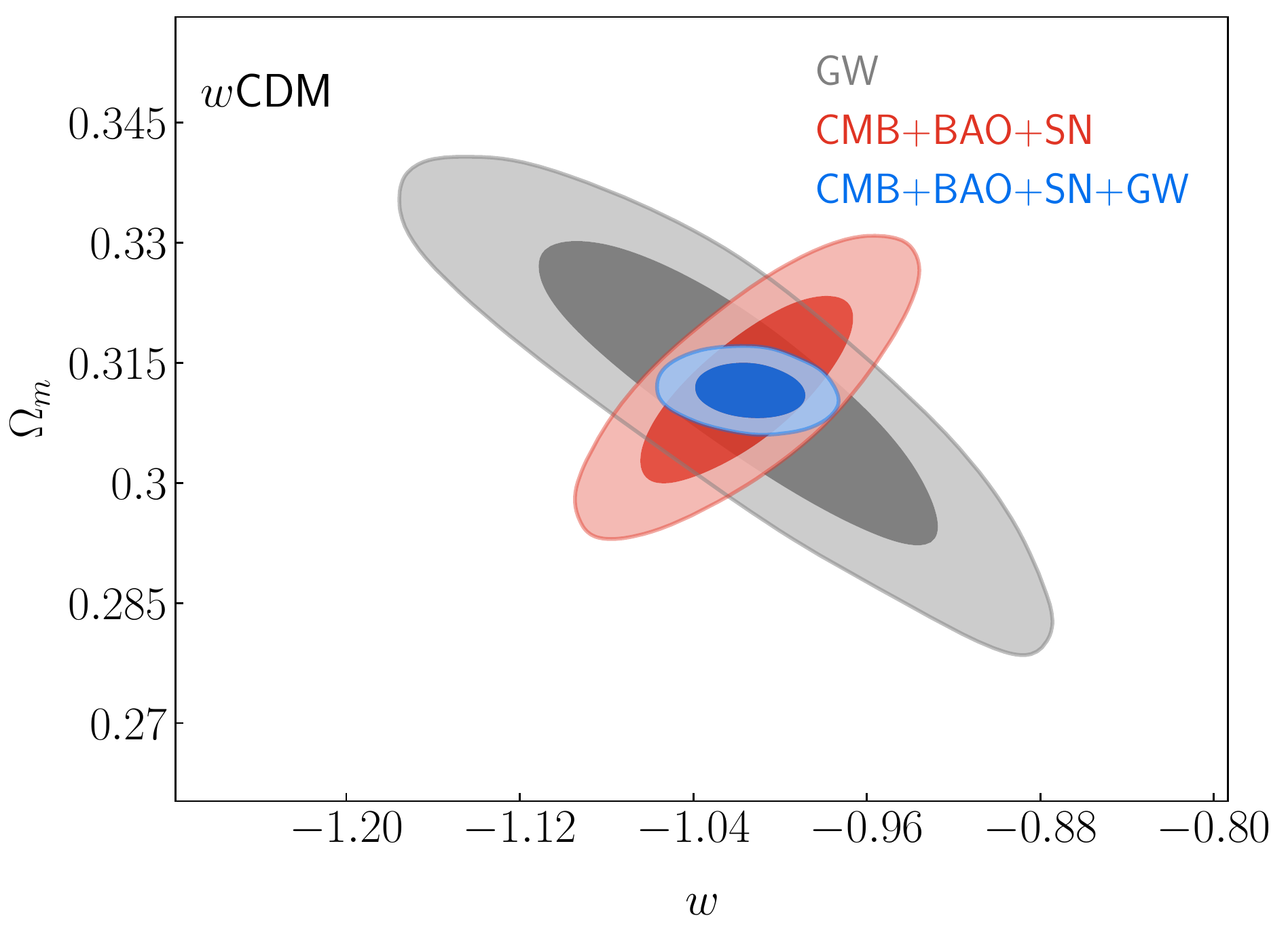}
\includegraphics[width=7.2cm,height=5.4cm]{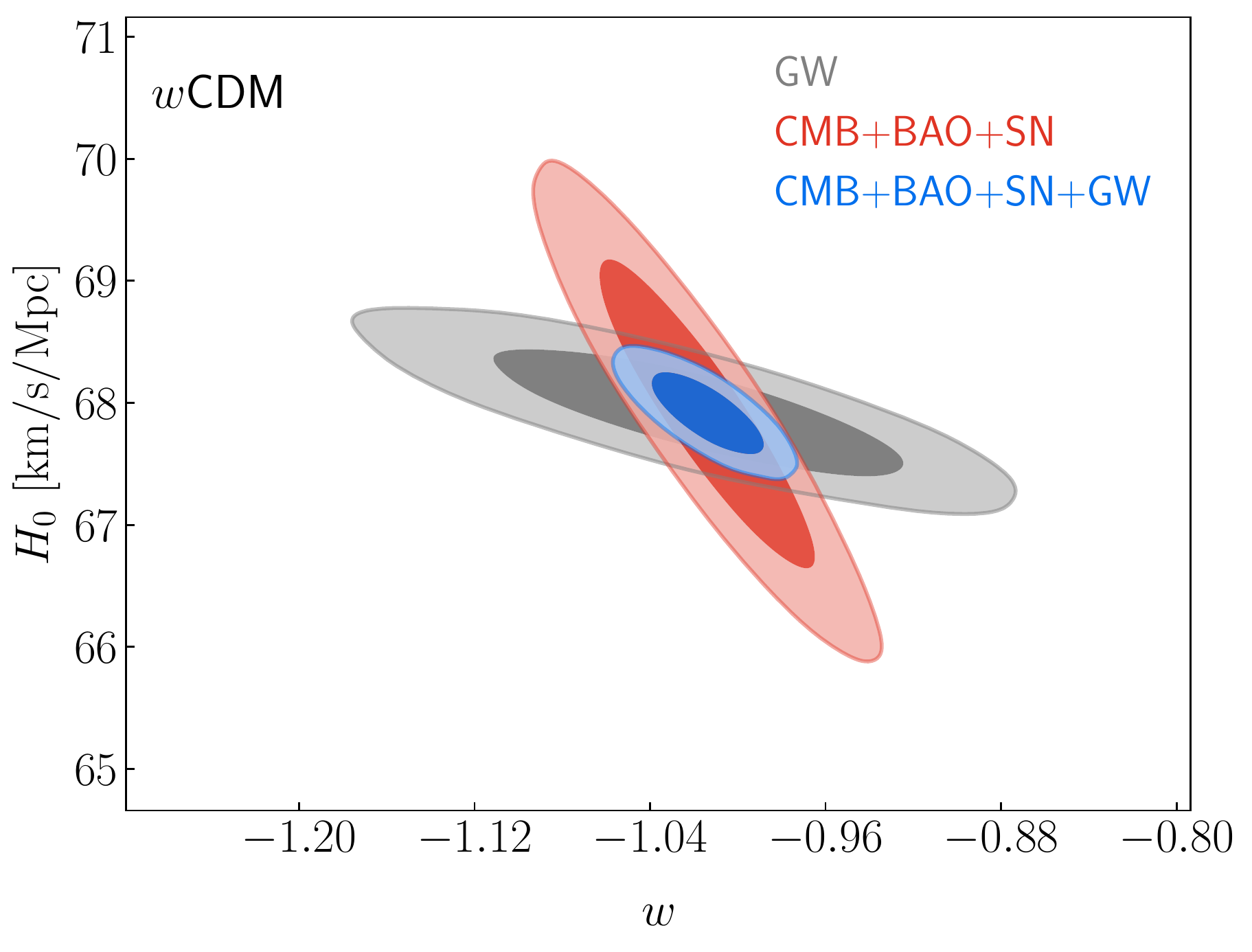}
\includegraphics[width=7.2cm,height=5.4cm]{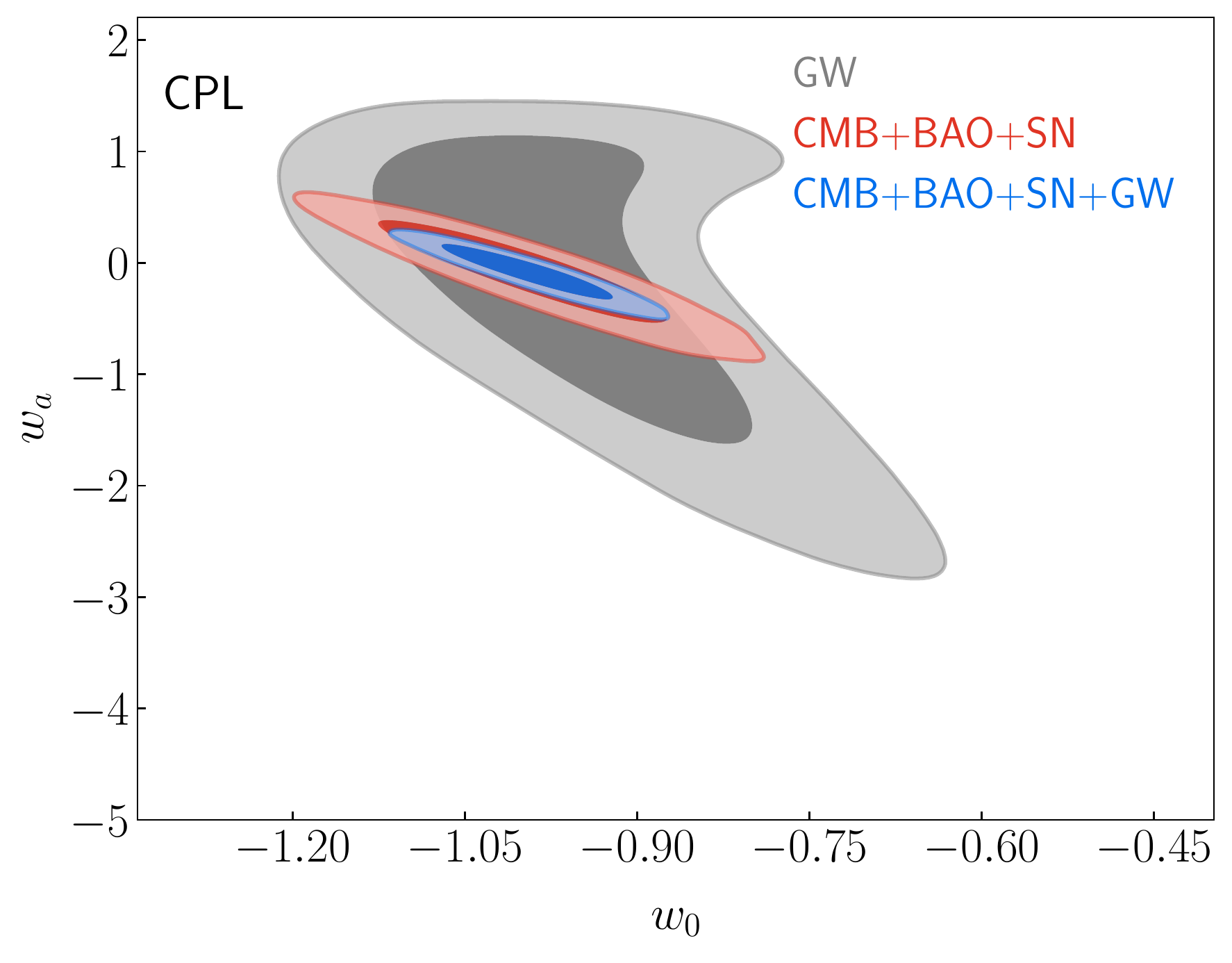}
\centering
 \caption{\label{fig:ce} Constraints (68.3\% and 95.4\% confidence level) on the $\Lambda$CDM, $w$CDM, and CPL models by using the GW, CMB+BAO+SN, and CMB+BAO+SN+GW data combinations.}
\end{figure*}

\begin{figure*}[!htbp]
\includegraphics[width=7.2cm,height=5.4cm]{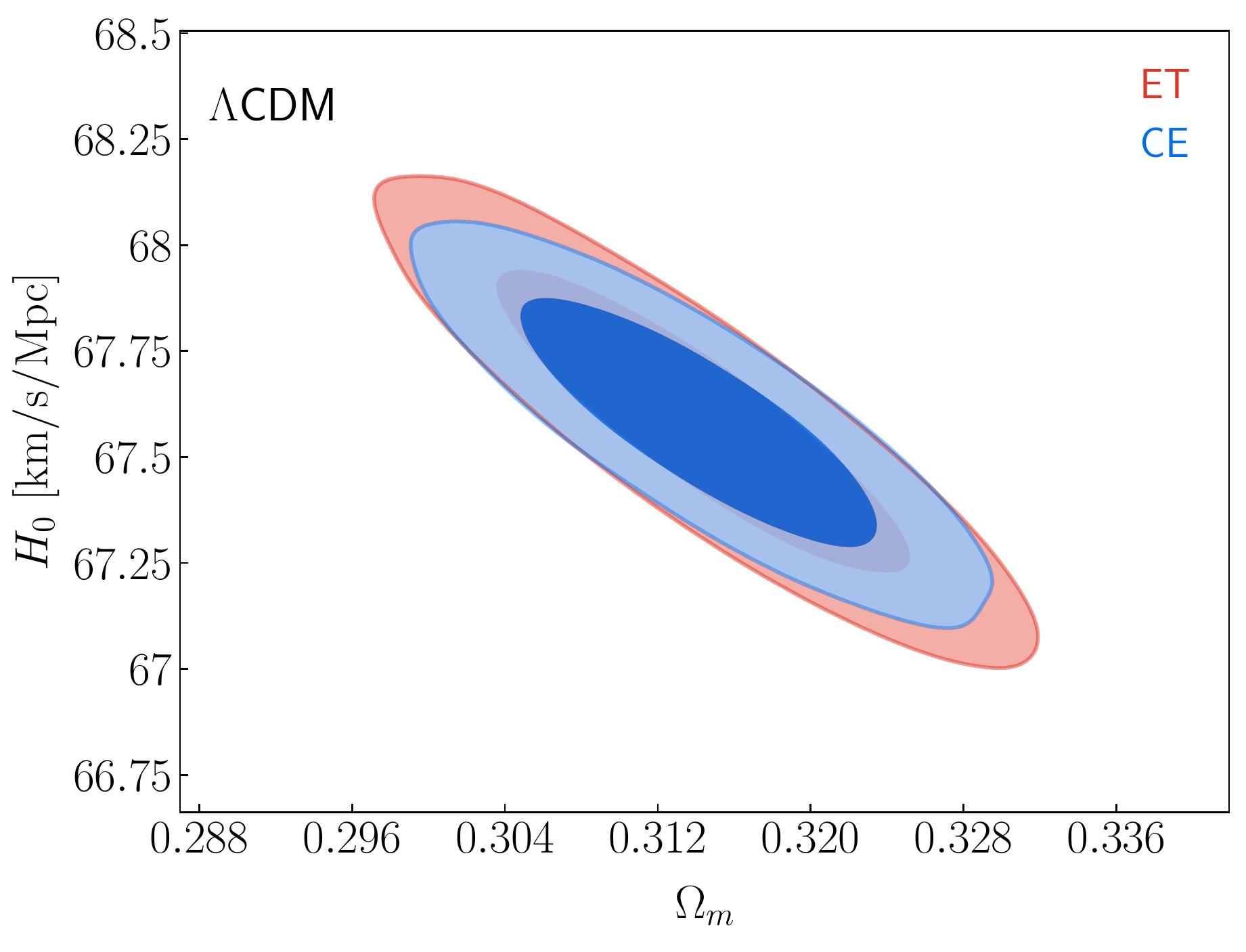}
\includegraphics[width=7.2cm,height=5.4cm]{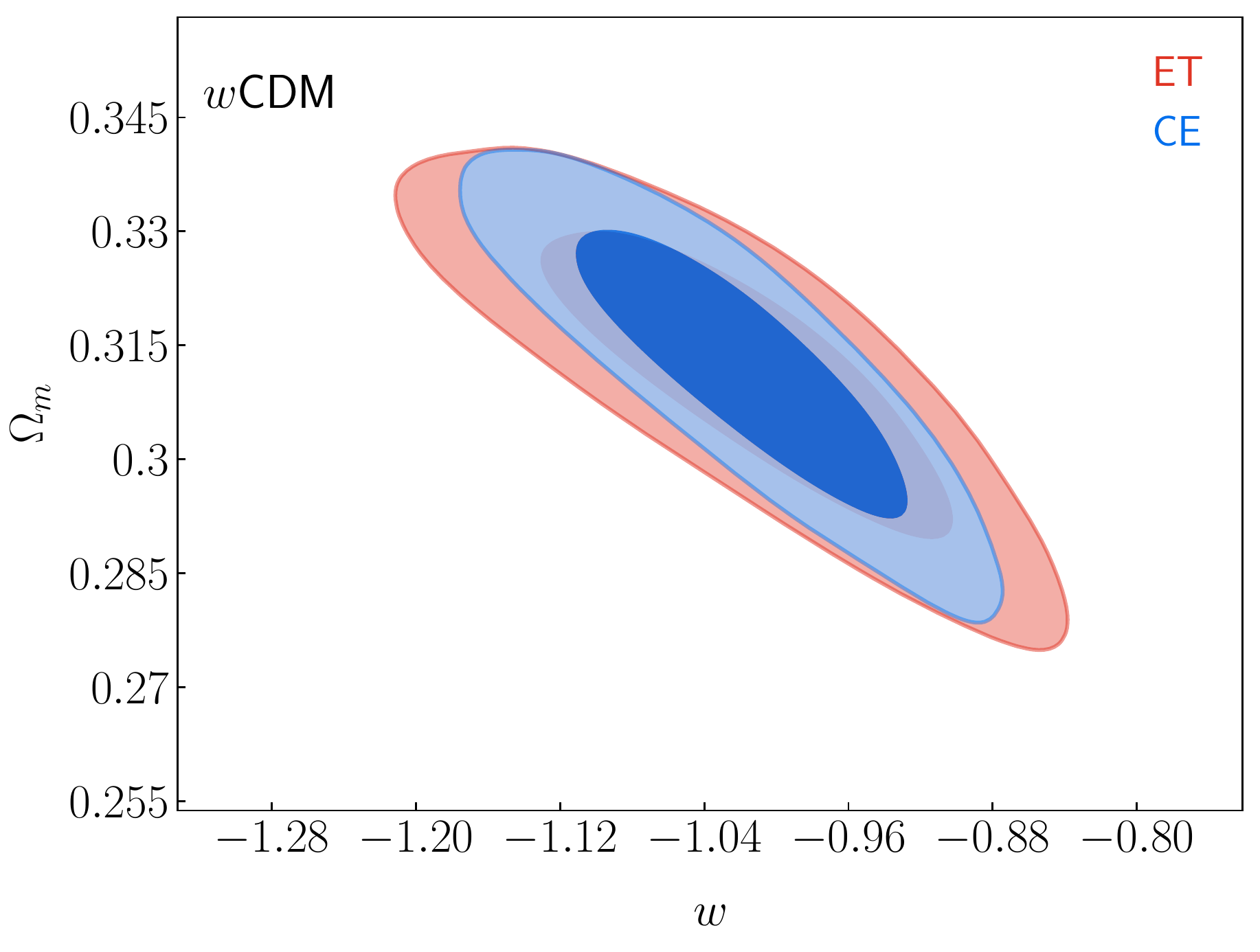}
\includegraphics[width=7.2cm,height=5.4cm]{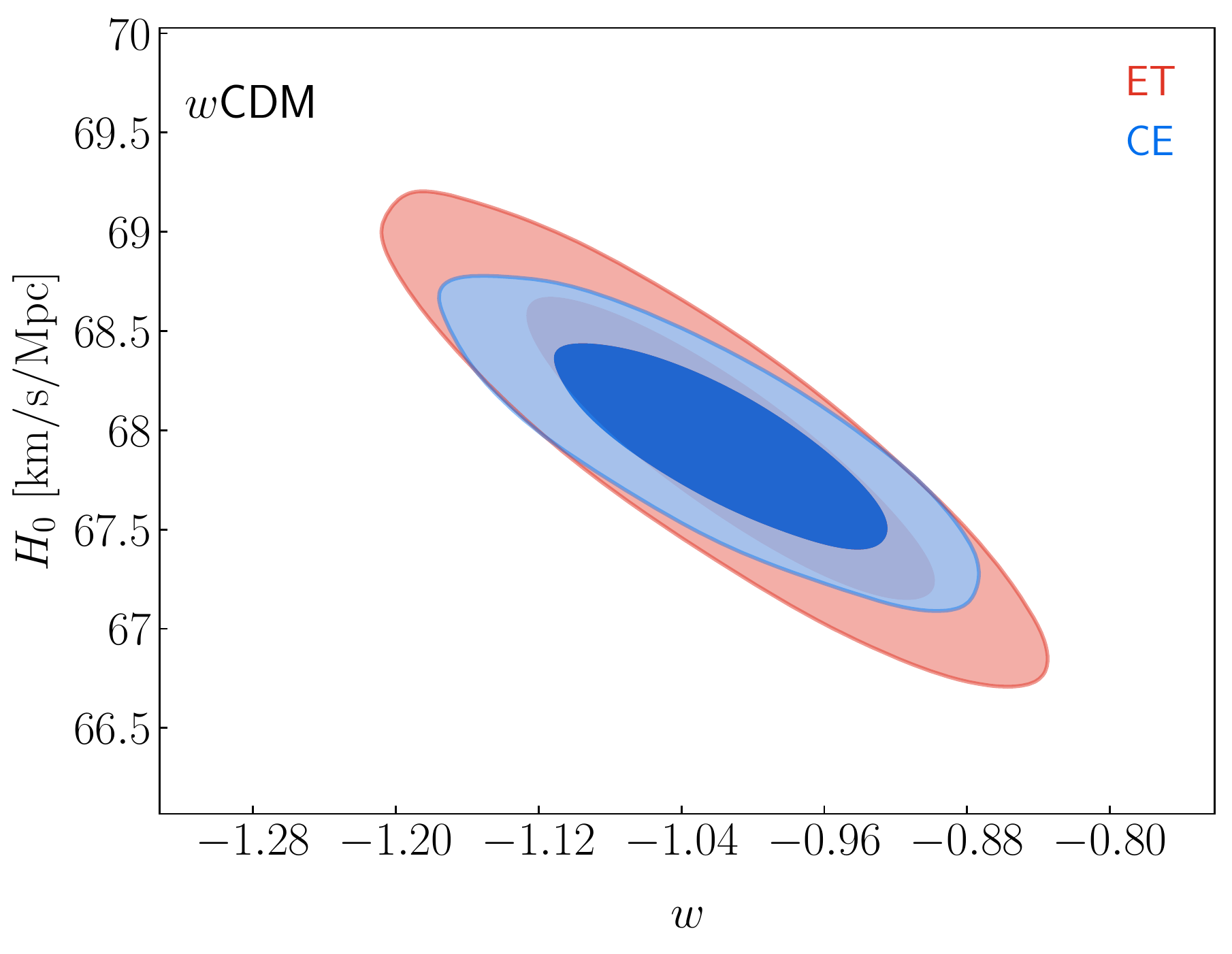}
\includegraphics[width=7.2cm,height=5.4cm]{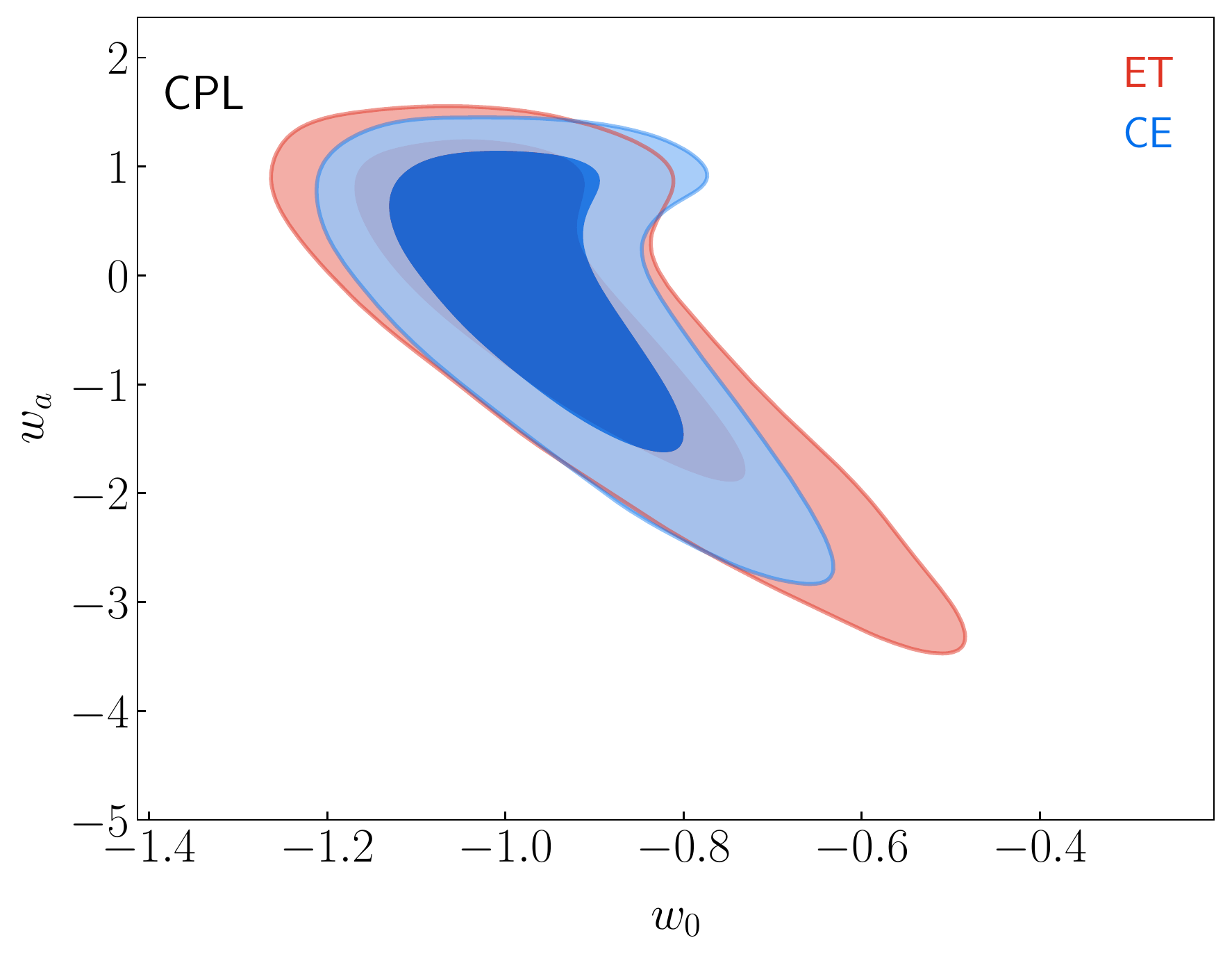}
\centering
 \caption{\label{fig:comparison} Constraints (68.3\% and 95.4\% confidence level) on the $\Lambda$CDM, $w$CDM, and CPL models by using the simulated data of ET and CE.}
\end{figure*}

\begin{table*}[!htbp]
\small
\setlength\tabcolsep{5.0pt}
\renewcommand{\arraystretch}{1.5}
\centering
\begin{tabular}{cccccccccc}\\
\hline
&& \multicolumn{2}{c}{$\Lambda$CDM}&&\multicolumn{2}{c}{$w$CDM}&&\multicolumn{2}{c}{CPL}\\
 \cline{3-4}\cline{6-7}\cline{9-10}
  && ET&CE && ET & CE   &&ET&CE\\
\hline
$\sigma(\Omega_{\rm m})$&&0.0071&0.0062&&0.0130&0.0125&&0.0450&0.0445\cr
$\sigma(H_0)$&&0.240&0.190&&0.510&0.340&&0.940&0.665\cr
$\sigma(w)$&&$-$&$-$&&0.075&0.061&&$-$&$-$\cr
$\sigma(w_0)$&&$-$&$-$&&$-$&$-$&&0.135&0.1045\cr
$\sigma(w_a)$&&$-$&$-$&&$-$&$-$&&1.015&0.900\cr\hline
$\varepsilon(\Omega_{\rm m})$&&0.0226&0.0197&&0.0418&0.0400&&0.1485&0.1503\cr
$\varepsilon(H_0)$&&0.0036&0.0028&&0.0075&0.0050&&0.0139&0.0098\cr
$\varepsilon(w)$&&$-$&$-$&&0.0737&0.0598&&$-$&$-$\cr
$\varepsilon(w_0)$&&$-$&$-$&&$-$&$-$&&0.1436&0.1098\cr
\hline
\end{tabular}
\centering
\caption{\label{tab3} Constraint errors and accuracies for the cosmological parameters in the $\Lambda$CDM, $w$CDM and CPL models by using the simulated data of ET and CE.}
\end{table*}

\begin{figure*}[!htbp]
\includegraphics[width=7.2cm,height=5.4cm]{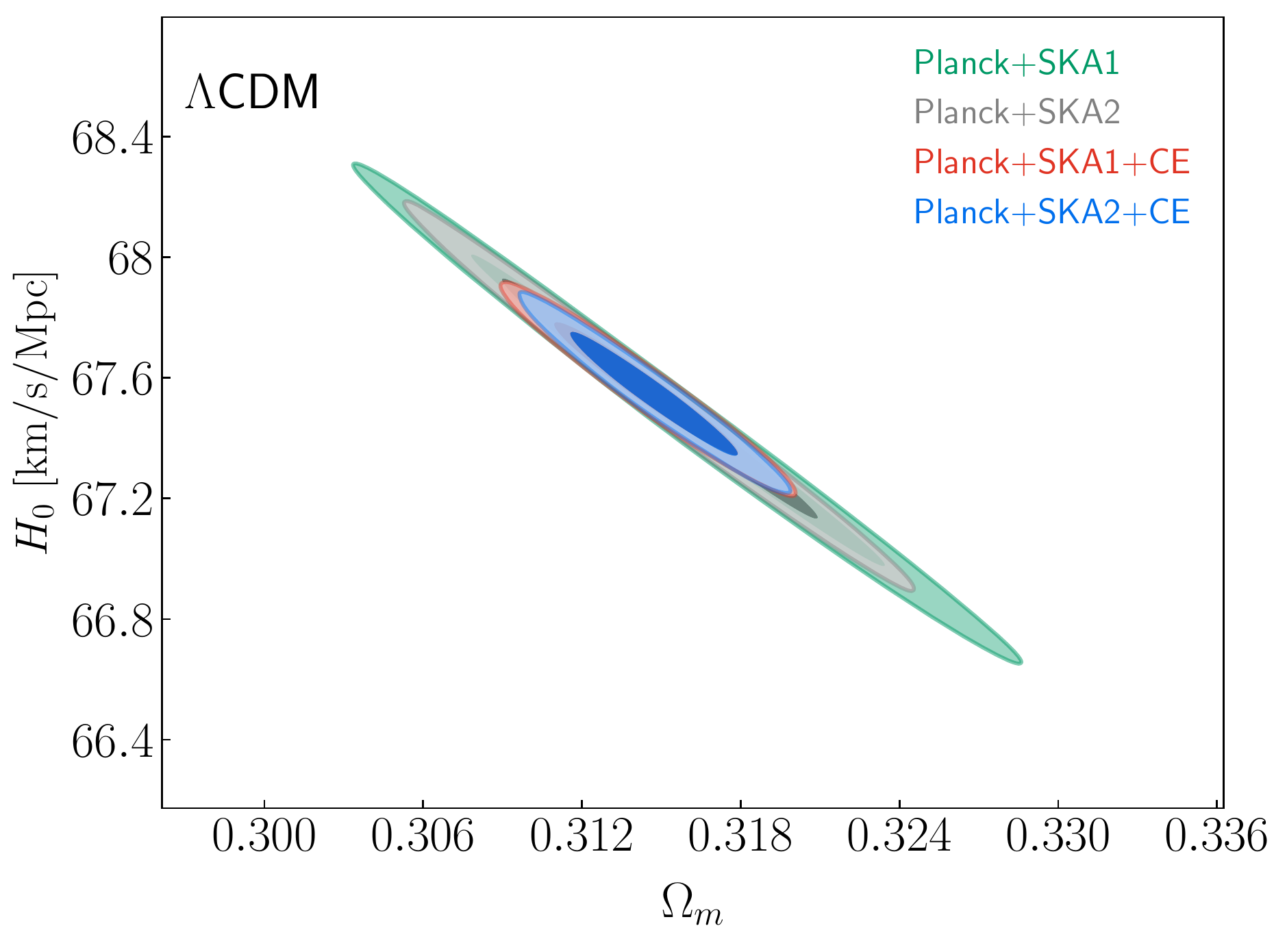}
\includegraphics[width=7.2cm,height=5.4cm]{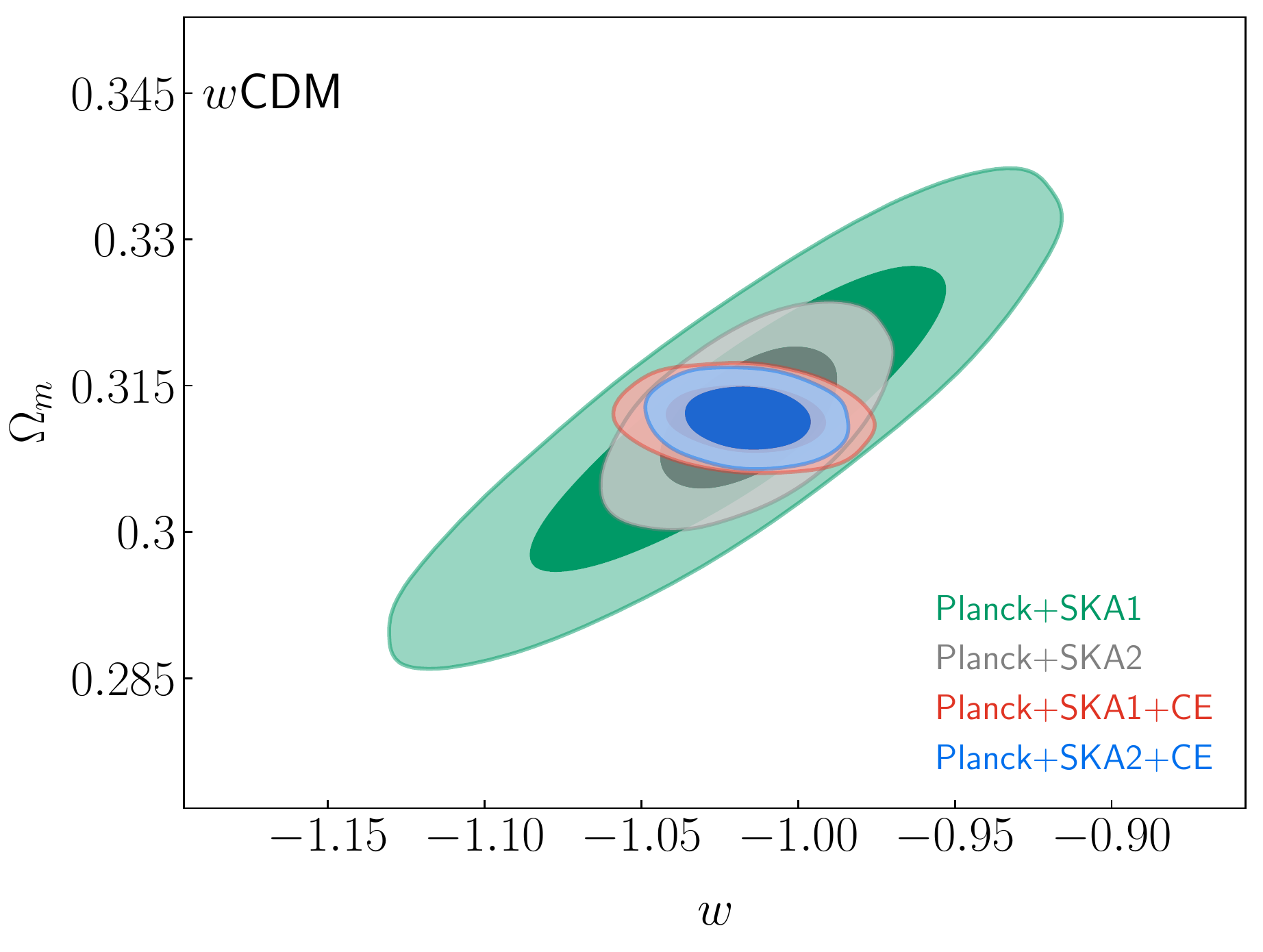}
\includegraphics[width=7.2cm,height=5.4cm]{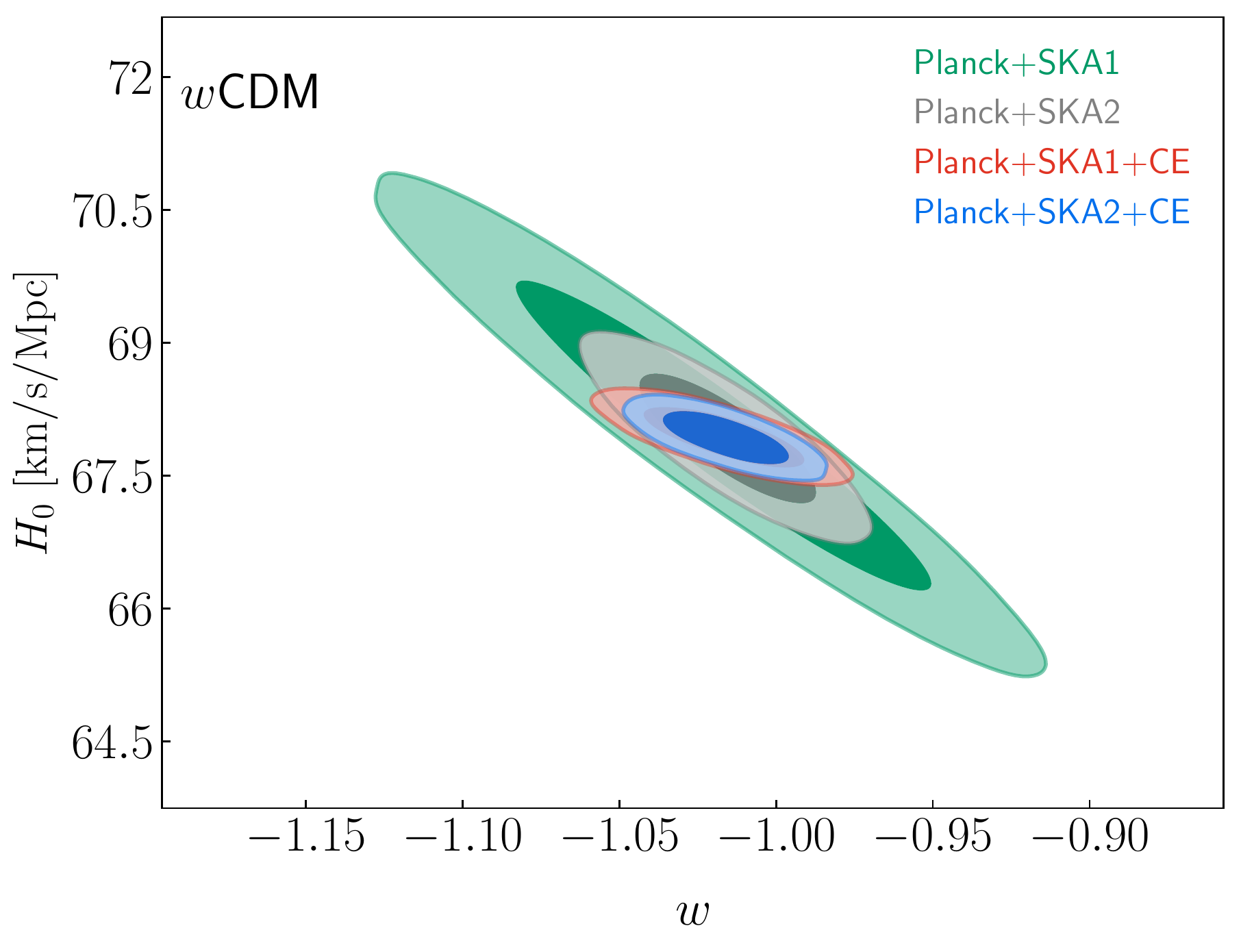}
\includegraphics[width=7.2cm,height=5.4cm]{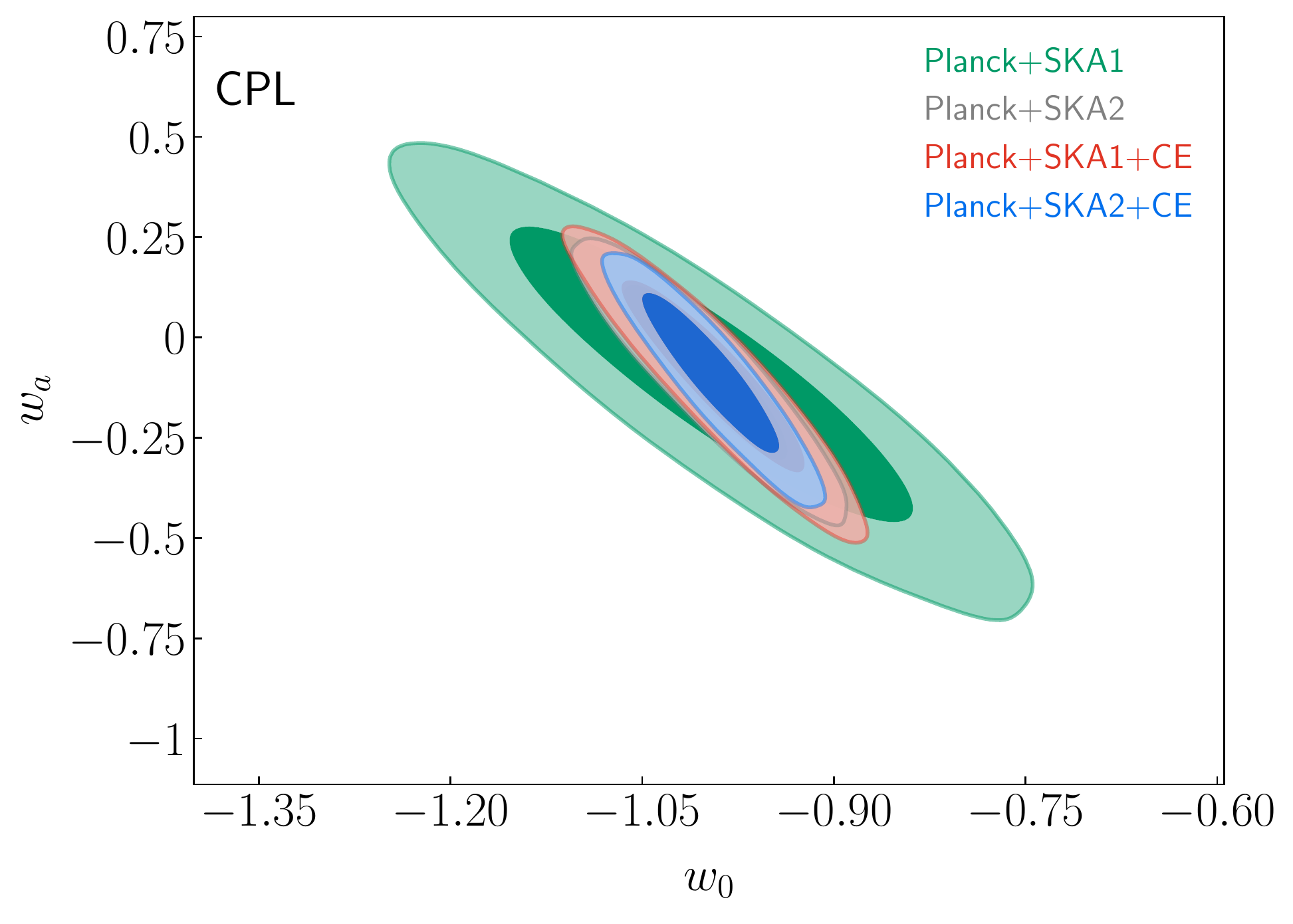}
\centering
 \caption{\label{fig:synergy} Constraints (68.3\% and 95.4\% confidence level) on the $\Lambda$CDM, $w$CDM, and CPL models by using the Planck+SKA1, Planck+SKA2, Planck+SKA1+CE, Planck+SKA2+CE data combinations.}
\end{figure*}

\begin{table*}[!htbp]
\renewcommand{\arraystretch}{1.5}
\begin{tabular}{cc c c c c}\\
\hline
Data&Planck+SKA1&Planck+SKA2&Planck+SKA1+CE&Planck+SKA2+CE\\
\hline
$\Omega_{\rm m}$&$0.3158^{+0.0052}_{-0.0052}$&$0.3149^{+0.0040}_{-0.0040}$&$0.3144^{+0.0023}_{-0.0023}$&$0.3147^{+0.0021}_{-0.0021}$\\
$H_0$&$67.48^{+0.34}_{-0.34}$&$67.53^{+0.26}_{-0.26}$&$67.56^{+0.15}_{-0.15}$&$67.55^{+0.14}_{-0.14}$\\
\hline
$\sigma(\Omega_{\rm m})$&$0.0052$&$0.0040$&$0.0023$&$0.0021$\\
$\sigma(H_0)$&$0.34$&$0.26$&$0.15$&$0.14$\\
\hline
$\varepsilon(\Omega_{\rm m})$&$0.0165$&$0.0127$&$0.0073$&$0.0067$\\
$\varepsilon(H_0)$&$0.0050$&$0.0039$&$0.0022$&$0.0021$\\
\hline
\end{tabular}
\caption{\label{tab5}The best-fit values of parameters, corresponding constraint errors and constraint accuracies within the $\Lambda$CDM model by using the Planck+SKA1, Planck+SKA2, Planck+SKA1+CE, and Planck+SKA2+CE data combinations. Here, $H_0$ is in units of ${\rm km\ s^{-1}\ Mpc^{-1}}$.}
\end{table*}

\begin{table*}[!htbp]
\renewcommand{\arraystretch}{1.5}
\begin{tabular}{cc c c c c}\\
\hline
Data&Planck+SKA1&Planck+SKA2&Planck+SKA1+CE&Planck+SKA2+CE\\
\hline
$\Omega_{\rm m}$&$0.312^{+0.010}_{-0.010}$&$0.3117^{+0.0048}_{-0.0048}$&$0.3116^{+0.0023}_{-0.0023}$&$0.3117^{+0.0021}_{-0.0021}$\\
$H_0$&$68.00^{+1.10}_{-1.20}$&$67.93^{+0.49}_{-0.49}$&$67.94^{+0.22}_{-0.22}$&$67.93^{+0.20}_{-0.20}$\\
$w$&$-1.019^{+0.046}_{-0.041}$&$-1.016^{+0.019}_{-0.019}$&$-1.017^{+0.017}_{-0.017}$&$-1.016^{+0.013}_{-0.013}$\\
\hline
$\sigma(\Omega_{\rm m})$&$0.0100$&$0.0048$&$0.0023$&$0.0021$\\
$\sigma(H_0)$&$1.15$&$0.49$&$0.22$&$0.20$\\
$\sigma(w)$&$0.0435$&$0.0190$&$0.0170$&$0.0130$\\
\hline
 $\varepsilon(\Omega_{\rm m})$&$0.0321$&$0.0154$&$0.0074$&$0.0067$\\
$\varepsilon(H_0)$&$0.0169$&$0.0072$&$0.0032$&$0.0029$\\
$\varepsilon(w)$&$0.0427$&$0.0187$&$0.0167$&$0.0128$\\
\hline
\end{tabular}
\caption{\label{tab6}Same as table \ref{tab5}, but for the $w$CDM model (adding one more parameter, $w$).}
\end{table*}

\begin{table*}[!htbp]
\renewcommand{\arraystretch}{1.5}
\begin{tabular}{cc c c c c}\\
\hline
Data&Planck+SKA1&Planck+SKA2&Planck+SKA1+CE&Planck+SKA2+CE\\
\hline
$\Omega_{\rm m}$&$0.312^{+0.014}_{-0.016}$&$0.3108^{+0.0066}_{-0.0066}$&$0.3121^{+0.0046}_{-0.0046}$&$0.3118^{+0.0042}_{-0.0042}$\\
$H_0$&$68.00^{+1.60}_{-1.60}$&$68.07^{+0.58}_{-0.58}$&$67.92^{+0.39}_{-0.39}$&$67.95^{+0.31}_{-0.31}$\\
$w_0$&$-0.99^{+0.10}_{-0.10}$&$-0.999^{+0.044}_{-0.044}$&$-0.992^{+0.048}_{-0.048}$&$-0.995^{+0.035}_{-0.035}$\\
$w_a$&$-0.10^{+0.24}_{-0.24}$&$-0.09^{+0.16}_{-0.13}$&$-0.10^{+0.17}_{-0.15}$&$-0.09^{+0.14}_{-0.12}$\\
\hline
$\sigma(\Omega_{\rm m})$&$0.0150$&$0.0066$&$0.0046$&$0.0042$\\
$\sigma(H_0)$&$1.60$&$0.58$&$0.39$&$0.31$\\
$\sigma(w_0)$&$0.100$&$0.044$&$0.048$&$0.035$\\
$\sigma(w_a)$&$0.240$&$0.145$&$0.160$&$0.130$\\
\hline
$\varepsilon(\Omega_{\rm m})$&$0.0481$&$0.0212$&$0.0147$&$0.0135$\\
$\varepsilon(H_0)$&$0.0235$&$0.0085$&$0.0057$&$0.0046$\\
$\varepsilon(w_0)$&$0.1010$&$0.0440$&$0.0484$&$0.0352$\\
\hline
\end{tabular}
\caption{\label{tab7}Same as table \ref{tab5}, but for the CPL model (adding two more parameters, $w_0$ and $w_a$).}
\end{table*}

\section{Results}\label{sec4}

In this section, we shall report the constraint results for the cosmological parameters in the $\Lambda$CDM, $w$CDM, and CPL models.
We constrain the considered cosmological models with GW, CMB+BAO+SN, and CMB+BAO+SN+GW data combinations to complete our analysis.  The constraint results are shown in figure~\ref{fig:ce} and summarized in tables~\ref{tab1}--\ref{tab2}.
In figure~\ref{fig:ce}, we display the two-dimensional posterior distribution contours for various model parameters constrained at 68\% and 95\% confidence level (C.L.).
In tables~\ref{tab1}--\ref{tab2}, we exhibit the best-fit values with 1$\sigma$ errors quoted and the constraint errors and accuracies for the concerned parameters (i.e., $\Omega_{\rm m}$, $H_0$, $w$, $w_0$, and $w_a$). Note that, for a parameter $\xi$, we use $\sigma(\xi)$ and $\varepsilon(\xi)$ to denote its absolute and relative errors in the cosmological fit, respectively.

At first glance of figure~\ref{fig:ce}, we can clearly find that the future GW observation from CE could significantly improve the constraints on almost all the parameters to some different extent; for more details, see also tables~\ref{tab1}--\ref{tab2}. Particularly, in the $\Omega_m-H_0$ plane for the $\Lambda$CDM model, we find that the addition of CE mock data to the current optical observations (i.e., the CBS datasets) could break the degeneracy between the matter density and the Hubble constant, and further improve the constraint accuracies in the cosmological fit to a great extent. Concretely, when adding the simulated CE data to the CBS datasets, the constraint precisions of $\Omega_m$ and $H_0$ in the $\Lambda$CDM model could be improved from 1.59\% and 0.49\% to 0.70\% and 0.21\%, respectively.
Moreover, as shown in the planes of $w-H_0$ and $w-\Omega_{\rm m}$, we also find that the parameter degeneracy orientations of GW evidently differ from those of CBS data combination, which implies that the CE mock data can help to largely break the parameter degeneracies between these parameters in the $w$CDM model. With the help of CE mock data, the constraint precisions of $\Omega_m$, $H_0$ and $w$ can be improved from 2.47\%, 1.22\%, and 3.15\% to 0.74\%, 0.32\%, and 1.68\%, respectively.
Furthermore, the constraint results on the CPL model are shown in the  $w_0-w_a$ plane. We can apparently find that the constraining capability of CE is powerful as well, and the constraint accuracies of $\Omega_m$, $H_0$ and $w_0$ could be improved from 2.50\%, 1.24\%, and 8.37\% to 1.38\%, 0.54\%, and 5.04\%, respectively, with the addition of CE mock data. What is supposed to be emphasized is that the central value of $w_a$ is around zero, and thus the absolute error for this parameter is more reliable for quantifying the improvement as the relative error would be impacted by its statistic fluctuations. For the parameter $w_a$, the absolute constraint error is improved from 0.305 to 0.160. Therefore, we can conclude that the GW standard siren observations from CE will be able to significantly improve the cosmological parameter constraints in the near future.

\section{Some discussions}\label{sec5}

\subsection{Comparison of CE and ET}
As a newly proposed third-generation GW detector, CE is a L-shaped single interferometer, with the angle between two arms equal to $90^\circ$ and the length of arms reaching an astonishing 40 km. While the same-type ground-based GW detector ET consists of three identical interferometers with 10 km arms, with the angle between two arms equal to $60^\circ$, forming an equilateral triangle.
Since both CE and ET belong to the third-generation ground-based GW detectors, it is necessary to make a comparison for the promotion effects on the parameter estimation from the GW standard siren observations of CE and ET. In figure~\ref{fig:etce}, we show the sensitivity curves of CE and ET. For the ET mock data, the simulation method is similar to that of CE as described in section~\ref{sec3:a}. Then, we use the CE and ET mock data to constrain the three considered DE models separately. The constraint results are shown in figure~\ref{fig:comparison} and summarized in table~\ref{tab3}.

From figure~\ref{fig:comparison}, we can clearly see that the constraint results from CE are slightly better than those from ET. Concretely, the {constraint} precisions of the parameters $\Omega_m$ and $H_0$ from CE are higher than those of ET by 1.1\%$-$12.7\%  and 20.8\%$-$33.3\%, respectively.  As for the EoS parameters $w$ and $w_0$, the {constraint} precisions from CE are higher than those of ET by 18.7\% and 22.6\%. For the parameter $w_a$, the constraint absolute error from CE is smaller than that of ET by 11.3\%. Thus, it can be seen that the constraining power of CE on the cosmological parameters maybe stronger than that of ET in the future.

\subsection{Synergy with SKA}
As an important non-optical cosmological probe besides the GW standard sirens, the HI 21 cm observation with the 
SKA will also play an important role in the future cosmological parameter estimation (see, e.g., refs.~\cite{Bacon:2018dui,Xu:2016kwz,Zhang:2019dyq,Zhang:2019ipd}). Hence, it is also necessary to discuss the synergy of the GW standard siren observation from the CE with the HI 21 cm radio observation from the SKA.
As a next-generation radio observatory, the SKA is currently under development, with the total collecting area being one square kilometer. Here, we consider the HI intensity mapping observations with the SKA phase 1 mid-frequency array (denoted as SKA1-MID), and the HI galaxy survey with SKA phase 2 (denoted as SKA2).
They are both able to observe the large-scale structure at the redshift range of $0\lesssim z\lesssim 3$ where DE dominates the evolution of the universe.

We use the simulated data of the BAO measurements from the HI sky survey based on SKA1 and SKA2 in ref.~\cite{Bull:2015nra}. For the SKA1-MID intensity mapping, the experimental specifications used in the forecast are given in table 2 of ref.~\cite{Bull:2015nra}. For the SKA2 HI galaxy survey, 
the expected galaxy number counts and bias of HI galaxies used in the forecast can be found in table 1 of ref.~\cite{Bull:2015nra}. The expected relative errors of $H(z)$ and $D_A(z)$ in the BAO measurements by the SKA, using the Fisher forecasting method, are given in figure 3 of ref.~\cite{Bull:2015nra}.

We employ the Planck+SKA1, Planck+SKA2, Planck+CE+SKA1, and Planck+CE+SKA2 data combinations to constrain the $\Lambda$CDM, $w$CDM, and CPL models and make the analysis. The constraint results are shown in figure~\ref{fig:synergy} and summarized in tables~\ref{tab5}--\ref{tab7}. In figure~\ref{fig:synergy}, we show the two-dimensional posterior distribution contours for the cosmological parameters (i.e., $\Omega_{\rm m}$, $H_0$, $w$, $w_0$, and $w_a$) constrained at 68\% and 95\% C.L.. In tables~\ref{tab5}--\ref{tab7}, we display the constraint errors and the constraint accuracies for the concerned parameters.

From figure~\ref{fig:synergy} and tables~\ref{tab5}--\ref{tab7}, we can apparently find that regardless of adding the CE mock data to either the Planck+SKA1 datasets or the Planck+SKA2 datasets, the constraint results could be significantly improved in all the considered DE models.
For example, when adding the CE mock data to the Planck+SKA1 datasets, the constraints on the matter density $\Omega_m$ can be improved by 55.8\%$-$77\%, and the constraints on the Hubble constant $H_0$ can be promoted by 55.9\%$-$80.9\%. With respect to the parameters featuring the property of DE, the improvements are also evident, with the constraints on the parameter $w$ in the $w$CDM model promoted by 60.9\%, and the parameters of $w_0$ and $w_a$ in the CPL model improved by 52\% and 33.3\%, respectively. Hence, we can conclude that the synergy between the GW standard siren observation from CE and the 21 cm emission observation from SKA would be extremely effective on the cosmological parameter estimation.

\section{Conclusion}\label{sec6}

In this paper, we have investigated how the future GW standard sirens observation from CE impact the cosmological parameter estimation. For the conventional cosmological probes based on the EM observation, we use the latest CMB data from Planck 2018, the optical BAO measurements, and the SN observation of Pantheon compilation. We consider three typical cosmological DE models in this work, including the $\Lambda$CDM model, the $w$CDM model and the CPL model. In order to quantify the constraining ability of the additional GW data, we consider another two data combinations, namely CBS and CBS+GW, to constrain these DE models.

We find that the GW standard siren observation from the CE could tremendously improve the constraints on the cosmological parameters (i.e., $\Omega_{\rm m}$, $H_0$, $w$, $w_0$, and $w_a$). With the addition of CE in CBS datasets, the constraints on $\Omega_m$ can be improved by 44.9\%$-$70.1\%, and the constraints on $H_0$ can be improved by 56.0\%$-$73.5\%. For the constraints on those EoS parameters, the improvements are also evident, with the parameter $w$ in $w$CDM model promoted by 46.9\%, $w_0$ and $w_a$ in CPL model promoted by 39.8\% and 47.5\% respectively. What's more, we also find that degeneracies between several cosmological parameters, such as $\Omega_m$ and $H_0$ in $\Lambda$CDM model, $\Omega_m$ and $w$ in $w$CDM model as well as $w_0$ and $w_a$ in CPL model, could be significantly broken with the addition of CE mock data in the cosmological fit. Therefore, we conclude that the future GW standard siren from the CE could provide significant improvement in parameter estimation.

In addition, we make a comparison for the parameter constraining capability of the GW standard siren observations from ET and CE. We find that the constraint results from CE are slightly better than those from ET. Furthermore, we discuss the synergy of the GW standard siren observation from the CE with the HI 21 cm radio waves observation from the SKA. We find that the synergy between the GW standard siren observation from CE and the 21 cm emission observation from SKA will be effective on the cosmological parameter estimation in the future.

\begin{acknowledgments}
 We thank Jing-Zhao Qi and Hai-Li Li for helpful discussions. This work was supported by the National Natural Science Foundation of China (Grant Nos. 11975072, 11690021, 11875102, 11835009, 11973047, and 11633004), the CAS Strategic Priority Research Program (Grant No. XDA15020200), the Liaoning Revitalization Talents Program (Grant No. XLYC1905011), and the Fundamental Research Funds for the Central Universities (Grant No. N2005030).
\end{acknowledgments}

\end{document}